\pdfoutput=1
\documentclass[aps,prl,floatfix,superscriptaddress,twocolumn,reprint]{revtex4-1}
\usepackage{eurosym}
\usepackage{amsbsy}
\usepackage{latexsym,epsfig,graphicx}
\usepackage{dcolumn}
\usepackage{graphicx}
\usepackage{subfigure}
\usepackage{comment}
\usepackage{color}
\usepackage{bm}
\usepackage{mathrsfs}
\usepackage{amssymb}
\usepackage{amsfonts}
\usepackage{mathrsfs}
\usepackage{amsmath}
\usepackage{color}
\usepackage{graphicx}
\usepackage{bm}
\usepackage{amssymb}
\usepackage{xspace}
\usepackage{epstopdf}
\usepackage{dcolumn}
\usepackage{tabularx}
\usepackage{longtable}
\usepackage[colorlinks=true, letterpaper=true, pdfstartview=FitV, linkcolor=blue, citecolor=blue, urlcolor=blue]{hyperref}
\usepackage[normalem]{ulem}
\usepackage[version=4]{mhchem}

\newcommand{\sgn}{\text{sgn}}

\begin{document}
\title{Chiral to helical Majorana fermion transition in a $p$-wave superconductor}
\author{Haiping Hu}
\affiliation{Department of Physics and Astronomy, George Mason University, Fairfax, Virginia 22030, USA}
\affiliation{Department of Physics and Astronomy, University of Pittsburgh, Pittsburgh, Pennsylvania 15260, USA}
\author{Indubala I. Satija}
\affiliation{Department of Physics and Astronomy, George Mason University, Fairfax, Virginia 22030, USA}
\author{Erhai Zhao}
\affiliation{Department of Physics and Astronomy, George Mason University, Fairfax, Virginia 22030, USA}
\date{\today}
\begin{abstract}
Chiral and helical Majorana edge modes are two archetypal gapless excitations of two-dimensional topological superconductors. They belong to superconductors from two different Altland-Zirnbauer symmetry classes characterized by $\mathbb{Z}$ and $\mathbb{Z}_2$ topological invariant respectively. It seems improbable to tune a pair of co-propagating chiral edge modes to counter-propagate without symmetry breaking. Here we show that such a direct topological transition is in fact possible, provided the system possesses an additional symmetry $\mathcal{O}$ which changes the bulk topological invariant to $\mathbb{Z}\oplus \mathbb{Z}$ type. A simple model describing the proximity structure of a Chern insulator and a $p_x$-wave superconductor is proposed and solved analytically to illustrate the transition between two topologically nontrivial phases. The weak pairing phase has two chiral Majorana edge modes, while the strong pairing phase is characterized by $\mathcal{O}$-graded Chern number and hosts a pair of counter-propagating Majorana fermions. The bulk topological invariants and edge theory are worked out in detail. Implications of these results to topological quantum computing based on Majorana fermions are discussed.
\end{abstract}
\maketitle
{\color{blue}\textit{Introduction.}}
A defining feature of topological quantum matter \cite{review1,review2} is the existence of protected boundary/edge modes due to the nontrivial topology of the bulk material. Gapped bulk Hamiltonians can be classified into different Altland-Zirnbauer (AZ) classes based on their symmetries \cite{class1,class2,class3,class4}. Each symmetry class has its own manifestation of the bulk-boundary correspondence. A well known example is the two-dimensional (2D) quantum Hall insulator in class A, which is characterized by an integer Chern number and has gapless chiral edge modes. Another example is the quantum spin Hall insulator \cite{ti1,ti2,ti3} in class AII with counter-propagating (helical) edge modes protected by time-reversal symmetry, and characterized by a $\mathbb{Z}_2$ invariant. Analogously, in 2D topological superconductors, chiral Majorana edge modes appear in class-D superconductors such as $p_x+ip_y$ \cite{chiralsc1}, while helical Majorana modes emerge in class-DIII superconductors with time-reversal symmetry \cite{helical1,helical2,helical3,helical4,helical5,helical6,helical7,helical8}. These edge modes, commonly referred to as Majorana fermions, are dispersive and differ from spatially localized Majorana zero modes which obey anyonic statistics. Recently, however, it is shown that the propagation of chiral Majorana fermions can lead to the same unitary transformation as the braiding of Majorana zero modes \cite{topoquantumchiral}. This opens up a new route for topological quantum computing \cite{topoquantum1,topoquantum2,topoquantum3,topoquantum4,topoquantum5,topoquantum6,topoquantum7} based on electrical manipulation of chiral Majorana fermions on the mesoscopic scale.

\begin{figure}[t]
\centering
\includegraphics[width=3.35in]{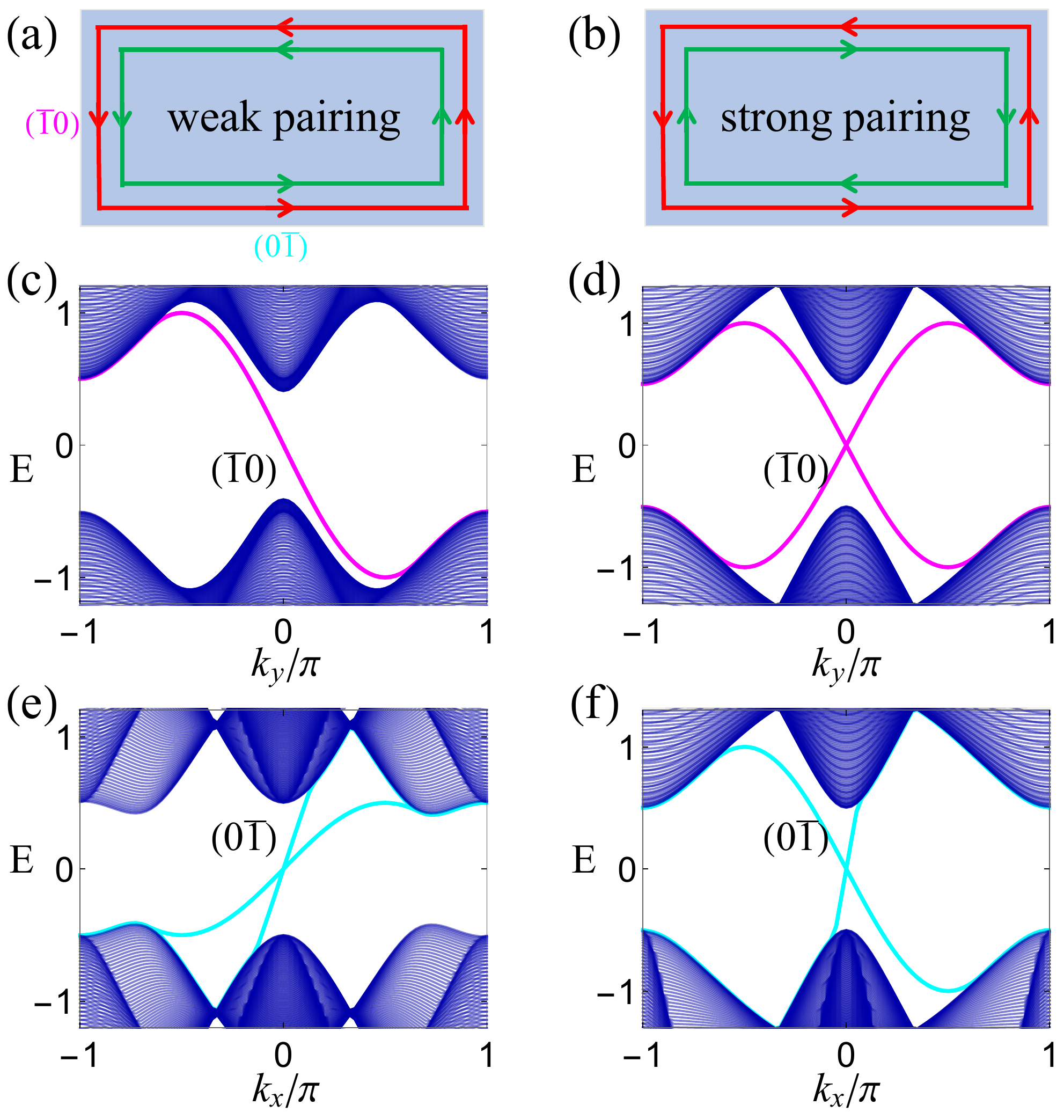}
\caption{Schematic of the chiral (a) and helical (b) Majorana edge modes in the weak and strong pairing phase of model (\ref{CIBdG}). The red/green line depicts edge modes in different subspaces of $\mathcal{O}$ symmetry, with the arrow indicating the propagating direction. Panels (c)-(f) show the quasiparticle spectra for the Hamiltonian Eq. (\ref{CIBdG}) in a half-infinite geometry with open boundary along a $(\bar{1}0)$ or $(0\bar{1})$ edge. Left column, (c) and (e): $\Delta=0.5$, the bulk is in the WP phase. Right column, (d) and (f): $\Delta=2$, the SP phase. $M=0.5$. The magenta and cyan lines label the Majorana edge modes on $(\bar{1}0)$ and $(0\bar{1})$ edges, respectively. In (c), the chiral edge modes are doubly degenerate. }
\label{edge}
\end{figure}

In this paper, we seek a tunable 2D superconductor that changes from the host of a pair of chiral Majorana modes to the host of a pair of counter-propagating Majorana modes, schematically shown in Fig. \ref{edge}(a) and (b), as the magnitude of the superconducting pairing is increased. At first sight, such a direct $\mathbb{Z}$ to $\mathbb{Z}_2$ transition seems impossible without any symmetry breaking because these two types of edge modes require distinct symmetries of the bulk Hamiltonian according to the standard AZ classification \cite{class1,class2,class3,class4}. However, as we show explicitly below, the presence of additional symmetry (besides time reversal, particle-hole and chiral symmetry) gives rise to richer physics beyond the AZ classes, and it is indeed possible to realize both types of edge modes within the same material. This implies that the direction of the Majorana fermions can be controlled to yield new quantum gates. Furthermore, we show that helical Majorana fermions, in addition to their chiral cousins, can also be used to achieve braiding-like operations. This significantly broadens the choice of topological superconductors for topological quantum computation.

Central to our proposal and analysis is a symmetry operator $\mathcal{O}$ that commutes with the Hamiltonian $H$, $[\mathcal{O},H]=0$. Within the pair of edge modes, indicated by the red and green arrows in Fig. \ref{edge}(a) and (b), each belongs to a specific sub-eigenspace of $\mathcal{O}$ and one of them changes its chirality across the topological phase transition. In the weak pairing (WP) phase, the two Majorana modes belonging to different $\mathcal{O}$-subspace propagate unidirectionally. In contrast, in the strong pairing (SP) phase with large pairing amplitude, the two Majorana modes move in opposite directions. We construct a new bulk topological invariant to characterize both the WP and SP phases and demonstrate the bulk-edge correspondence by working out the effective edge theory for both phases. The edge modes are stable against any perturbations that preserve the $\mathcal{O}$ symmetry.

{\color{blue}\textit{Model.}} First consider a simple model of Chern insulator with two orbitals, referred to as pseudospin $\uparrow$ and $\downarrow$, per site on a square lattice. The Hamiltonian is $H^0_{\bm k}=\alpha_x \sin k_x\sigma_x+\alpha_y \sin k_y\sigma_y+[M-t_0(\cos k_x+\cos k_y)]\sigma_z$. Here $t_0=1$ is the intra-orbital hopping amplitude and sets the energy unit, $\alpha_{x/y}$ is the inter-orbital hopping (analogous to spin-orbit coupling) along the $x/y$ direction, $M$ denotes an effective Zeeman field, and the Pauli matrices $\sigma_i$ are defined in orbital (pseudospin) space. This model is introduced for example in Refs. \cite{qah1,qah2} and recently realized using cold atoms in optical lattices \cite{qahexp1,qahexp2}. The topological properties of $H^0_{\bm k}$ is well known and characterized by Chern number $\mathcal{C}$. For $0<|M|<2 t_0$, $\mathcal{C}=\sgn(M)$ is simply given by the sign of $M$; while in other parameter regions, $\mathcal{C}=0$ and the system is a trivial band insulator. Now let us introduce on top of $H^0_{\bm k}$ an effective $p_x$-wave pairing term similar to the one in the Kitaev chain \cite{pwavechain}. Fermions of the same pseudospin at neighboring sites along the $x$ direction form Cooper pairs with amplitude $\Delta$ \cite{footnote}. The Bogoliubov de Gennes Hamiltonian $H=\frac{1}{2}\sum_{\bm{k}}\Psi^{\dag}_{\bm k}H_{\bm k}\Psi_{\bm k}$ in Nambu basis $\Psi^{\dag}_{\bm k}=(c_{\uparrow\bm k}^{\dag},c_{\downarrow \bm k}^{\dag},c_{\uparrow -\bm k},c_{\downarrow -\bm k})$ reads
\begin{eqnarray}
H_{\bm k}&=&\alpha_x\sin k_x \sigma_x+\alpha_y\sin k_y \sigma_y\tau_z\nonumber\\
&+&[M-t_0(\cos k_x+\cos k_y)]\sigma_z\tau_z+\Delta\sin k_x \tau_y,\label{CIBdG}
\end{eqnarray}
where the Pauli matrices $\tau_i$ operate in the particle-hole space. Like all class-D Hamiltonians in the AZ classification table, $H_{\bm k}$ has particle-hole symmetry $\mathcal{P}H_{\bm k}\mathcal{P}^{-1}=-H_{-\bm k}$ with $\mathcal{P}=\tau_x K$ and $K$ is the complex conjugate \cite{class1,class2,class3,class4}. In addition, there exists another important symmetry $\mathcal{O}=\sigma_x\tau_y$, which satisfies
\begin{eqnarray}
[\mathcal{O},H_{\bm k}]=0,~~~[\mathcal{O},\mathcal{P}]=0.
\end{eqnarray}
We emphasize that $\mathcal{O}$ involves both the pseudospin and particle-hole space. It is helpful to visualize the system as a bilayer, with each orbital confined to one of the two layers slightly shifted away from each other. Then $\mathcal{O}$ exchanges the two layers and generalizes the 2D mirror reflection \cite{mirrorsc}. It plays a crucial role in our analysis. The quasiparticle energy spectrum of $H_{\bm k}$ is given by
\begin{eqnarray}
\pm E_{\bm k}&=&\big[(\alpha_x\pm \Delta)^2\sin^2 k_x+\alpha_y^2\sin^2 k_y\notag\\
&+&(M-t_0\cos k_x-t_0\cos k_y)^2\big]^{1/2}. \label{spect}
\end{eqnarray}
We will focus on the parameter region $0<M<2 t_0$ to ensure the base insulator is topologically non-trivial. From Eq. \eqref{spect}, one observes with increasing $\Delta$, the gap closes when $\Delta=\alpha_x$ for a fixed $M$. The gap closing occurs at two Dirac points $\bm {k}_{\pm}\equiv(\pm k_x^0,k_y^0)=(\pm\arccos(\frac{M}{t_0}-1),0)$. We will refer to the gapped superconducting phase at $\Delta<\alpha_x$ as the WP phase; and the phase at $\Delta>\alpha_x$ as the SP phase.

Both phases turn out to be topologically nontrivial and give rise to Majorana edge states. Fig. \ref{edge} illustrates the quasiparticle spectra of $H_{\bm k}$ in a half-infinite geometry with $(\bar{1}0)$ edge (along $y$-direction), or $(0\bar{1})$ edge (along $x$-direction). For the WP phase [Fig. \ref{edge}(c)(e), left column], a pair of chiral Majorana modes propagate in the same direction on each edge. In particular, for the $(\bar{1}0)$ edge, the two modes are completely degenerate and they cross zero energy exactly at $k_x=0$. At the phase transition point $\Delta=\alpha_x$, the bulk bands touch at the two Dirac points, triggering a dramatic change in the edge spectrum. Within the SP phase [Fig. \ref{edge}(d)(f), right column], a pair of Majorana modes still cross zero energy at $\bm k=0$, but they have opposite group velocity and propagate in opposite directions along each edge.

{\color{blue}\textit{Topological invariant.}} According to the bulk-boundary correspondence, the emergence of edge states is governed by a nontrivial topological invariant in the bulk. For 2D class-D superconductors, the topological invariant is the Chern number defined as
\begin{eqnarray}
\mathcal{C}=\sum_{E_n<0}\int d^2 \bm k~\mathcal{F}^n_{k_x,k_y},
\end{eqnarray}
where $\mathcal{F}^n_{k_x,k_y}$ is the Berry curvature \cite{berry1,berry2} for the $n$-th band, and the sum is over all the occupied bands (with negative quasienergy). Note that for our model, the two lower bands overlap, $\mathcal{C}$ has to be calculated with care \cite{chern1,chern2}. We find for the WP phase, $\mathcal{C}=2$, in agreement with the number of chiral Majorana edge modes. This phase is adiabatically connected to a Chern insulator at $\Delta=0$, which hosts chiral, complex fermion as edge state. Such a chiral fermion can be decomposed into two chiral Majorana femions. Accordingly, one observes two gapless chiral modes on each edge in the WP phase. For the SP phase, the Chern number vanishes, $\mathcal{C}=0$. The change in $\mathcal{C}$ at the topological phase transition can be understood by expanding $H_{\bm k}$ around the two Dirac points. This leads to the following two-band Dirac Hamiltonian describing the gap closing:
\begin{eqnarray*}
H_{D}^{\bm k_{\pm}}(\delta \bm k)=\delta k_y\alpha_y\sigma_y\pm \delta k_x t_0\sin k_x^0\sigma_z\pm \delta m\sin k_x^0\sigma_x.
\end{eqnarray*}
Here $\delta m=\alpha_x-\Delta$ plays the role of Dirac mass. As the gap closes and reopens, $\delta m$ changes sign, accompanied by the change of Chern number $\Delta \mathcal{C}_{\bm k_{\pm}}=\frac{1}{2} [\sgn(\delta m<0)-\sgn (\delta m>0)]=-1$ through each Dirac point. Hence $\mathcal{C}$ changes by $-2$, yielding $\mathcal{C}=0$ for the SP phase, consistent with a pair of counter-propagating Majorana modes with no net chirality. Obviously $\mathcal{C}$ by itself cannot characterize the topology of the SP phase, or predict the number of Majorana edge modes in each direction.

We need to construct a new topological invariant. This can be accomplished by taking into account the symmetry $\mathcal{O}$ and following the general procedure of $K$-theory \cite{kgroup}. In the presence of $\mathcal{O}$, the topological invariant takes the form of $\mathbb{Z}\oplus\mathbb{Z}$, i.e., a pair of integers \cite{symmetryclass}. More physically, as $[\mathcal{O},H_{\bm k}]=0$, we can decompose $H_{\bm k}$ into two sectors. Each sector belongs to a specific sub-eigenspace of $\mathcal{O}$, labeled by $\pm1$, the eigenvalues of $\mathcal{O}$. For example, the transformation $V=e^{i\frac{\pi}{4}}I/2+e^{-i\frac{\pi}{4}}(\sigma_x\tau_x+\sigma_x\tau_y-\tau_z)/2$, which diagonalizes $\mathcal{O}$ via $V^{\dag}\mathcal{O}V=-\tau_z$, will take $H(\bm k)$ to $V^{\dag}H(\bm k)V=H_{-}(\bm k)\oplus H_{+}(\bm k)$, with
\begin{eqnarray}
H_{\pm}(\bm k)=&&\mp[M-t_0(\cos k_x+\cos k_y)]\sigma_z\notag\\
&&+(\alpha_x\pm\Delta)\sin k_x\sigma_x\mp\alpha_y\sin k_y\sigma_y.
\end{eqnarray}
As $[\mathcal{O},\mathcal{P}]=0$, $H_{\pm}(\bm k)$ each belongs to class $D$. Therefore, we can define the corresponding Chern number $\mathcal{C}_{\pm}$ in each $\mathcal{O}$-subspace.
The total Chern number discussed above is simply $\mathcal{C}=\mathcal{C}_{+}+\mathcal{C}_{-}$, while the $\mathcal{O}$-graded Chern number is defined as
\begin{eqnarray}
\mathcal{C}_{\mathcal{O}}=\frac{\mathcal{C}_+-\mathcal{C}_-}{2}.
\end{eqnarray}
This new topological invariant $\mathcal{C}_{\mathcal{O}}$ counts how many {\it pairs} of counter-propagating Majorana edge modes along each edge. For the WP phase, $\mathcal{C}_+=\mathcal{C}_-=1$, $\mathcal{C}=2$ and $\mathcal{C}_{\mathcal{O}}=0$; while for the SP phase, $\mathcal{C}_+=-\mathcal{C}_-=1$, yielding $\mathcal{C}=0$ and $\mathcal{C}_{\mathcal{O}}=\mathcal{C}_+=1$. The topological invariants and bulk-edge correspondence are summarized in Table \ref{Table}.
\begin{table}[h!]
\centering
\newcommand\T{\rule{0pt}{2.5ex}}
\newcommand\B{\rule[-1.7ex]{0pt}{0pt}}
\centering
\begin{tabular}{lccccl}
\hline\hline
Condition\quad & $\mathcal{C}_{+}$ & $\mathcal{C}_{-}$ & $\mathcal{C}$ & $\mathcal{C}_{\mathcal{O}}$ & Edge modes \T\\[3pt]
\hline
$\Delta<\alpha_x$ (WP) & 1 & 1 & $2$ & $0$ & co-propagating  \T\\
$\Delta>\alpha_x$ (SP) & 1 & -1 & $0$ & $1$ & counter-propagating  \T\\
\hline\hline
\end{tabular}
\caption{\label{Table}
{Bulk-edge correspondence for the WP and SP topological superconductors described by model (\ref{CIBdG}).}}
\end{table}

It is clear from these discussions that each Majorana edge mode belongs to a specific $\mathcal{O}$-subspace. They are protected by $\mathcal{O}$ in both phases. As long as this symmetry is preserved, the hybridization between them is forbidden. In comparison, for the time-reversal symmetry protected topological superconductors discussed in Ref. \onlinecite{helical1}, the helical edge modes form a Kramers pair, and they will switch partners under the time-reversal symmetry.

{\color{blue}\textit{Edge theory.}}
The essential features of the Majorana edge modes, and their symmetry properties, can be understood from their low-energy theory. Quite generally, an edge mode crossing zero energy can be regarded as bound state formed at a domain wall of the Dirac mass. The BdG Hamiltonian near $\bm k=(0,0)$ can be expanded to the second order in $\bm k$. It consists of three terms,
\begin{eqnarray}
H_{\bm k}=H_c + H_{k_x} + H_{k_y}. \label{decomp}
\end{eqnarray}
Here $H_c=m\sigma_z\tau_z$ with $m=M-2t_0$ being the effective Dirac mass, $H_{k_x}=\alpha_x k_x\sigma_x+\frac{t_0}{2}k_x^2\sigma_z\tau_z+\Delta k_x\tau_y$, and $H_{k_y}=\alpha_y k_y\sigma_y\tau_z+\frac{t_0}{2}k_y^2\sigma_z\tau_z$. To simplify the algebra, we take $\alpha_x=\alpha_y=t_0=1$ and assume $\Delta>0$ below.

Let us first consider the $(\bar{1}0)$ edge. The domain wall is located at $x=0$ and described by a spatially varying $m(x)$ which changes sign at $x=0$, i.e., $m(x>0)<0$ is the topological superconductor while $m(x<0)>0$ is the vacuum. We can replace $k_x$ with $-i\partial_{x}$ and treat $H_{k_y}$ as a perturbation in Eq. \eqref{decomp}. A trial wave function for the zero-energy bound state at the domain wall is given by $\phi_x e^{-\xi x}e^{i k_y y}$, where $\phi_x$ is a four-component spinor and the real number $\xi>0$ is the inverse localization length. The eigenvalue problem reduces to $\det[H_{k_x}+H_c]=0$, which gives an equation for $\xi$, $(m-\frac{\xi^2}{2})^2=\xi^2(\Delta\pm1)^2$. It has different solutions for the two phases.

For the WP phase with $\Delta<1$, $\xi$ has four solutions,
\begin{eqnarray*}
&&\xi_{1,\pm}=(1-\Delta)\pm\sqrt{(\Delta-1)^2+2m},\\
&&\xi_{2,\pm}=(1+\Delta)\pm\sqrt{(\Delta+1)^2+2m}.
\end{eqnarray*}
And the corresponding spinors are $\phi_{x,1}=(i,1,-i,1)^T/2$ and $\phi_{x,2}=(-i,,-1,-i,1)^T/2$, which transform as $\mathcal{O}\phi_{x,1}=-\phi_{x,1}$, $\mathcal{O}\phi_{x,2}=\phi_{x,2}$.
The superposition of these independent solutions have to satisfy the boundary condition $\Psi(x=0)=\Psi(x=+\infty)=0$.
This leads to the wave functions of the zero-energy bound states:
\begin{eqnarray}
&&\Psi_1(x)=C_1\phi_{x,1} (e^{-\xi_{1+} x}-e^{-\xi_{1-} x})e^{i k_y y},\\
&&\Psi_2(x)=C_2\phi_{x,2} (e^{-\xi_{2+} x}-e^{-\xi_{2-} x})e^{i k_y y}.
\end{eqnarray}
Here $C_1$ and $C_2$ are the normalization factors. The perturbation $H_{k_y}$ within the Hilbert space spanned by $\Psi_1(x)$ and $\Psi_2(x)$ is proportional to the identity matrix, $\langle\Psi_i(x)|H_{k_y}|\Psi_j(x)\rangle=-\alpha_y k_y\delta_{ij}$. Thus there are two degenerate chiral Majorana modes along the $(\bar{1}0)$ edge with group velocity $v_F=-\alpha_y$, in agreement with $\mathcal{C}=2$.

For the SP phase with $\Delta>1$, $\xi$ again has four solutions $\xi_{3,\pm}=(\Delta-1)\pm\sqrt{(\Delta-1)^2+2m}$, and $\xi_{4,\pm}=\xi_{2,\pm}$, with spinor $\phi_{x,3}=(i,-1,i,1)^T/2$ and $\phi_{x,4}=\phi_{x,2}$, respectively. And we observe that $\mathcal{O}\phi_{x,3}=-\phi_{x,3}$. The wave functions of the bound states are
\begin{eqnarray}
&&\Psi_3(x)=C_3\phi_{x,3} (e^{-\xi_{3+} x}-e^{-\xi_{3-} x})e^{i k_y y},\\
&&\Psi_4(x)=\Psi_2(x).
\end{eqnarray}
The perturbation $H_{k_y}$ in the Hilbert space spanned by $\Psi_3(x)$ and $\Psi_4(x)$ is found to be $\alpha_y k_y\sigma_z$. Therefore, there exist a pair of counter-propagating edge modes with group velocity $v_F=\pm \alpha_y$, in accordance with $\mathcal{C}=0$ and $\mathcal{C}_{\mathcal{O}}=1$. Note that the resulting edge theory is analogous to that of the quantum spin Hall system \cite{wucongjun}. Moreover, one of the edge modes, described by $\Psi_2(x)$ and $\Psi_4(x)$, smoothly evolves across the topological phase transition, while the other mode reverses the direction of travel.

Similar analysis can be performed for the $(0\bar{1})$ edge by treating $H_{k_x}$ as perturbation. With a trial wave function $\phi_y e^{-\xi y}e^{i k_x x}$, the existence of zero-energy bound states requires $\frac{\xi^2}{2}-\xi-m=0$, yielding two solutions $\xi_{\pm}=1\pm\sqrt{1+2m}$ and corresponding spinor $\phi_{y,1}=(i,i,1,1)^T/2$, $\phi_{y,2}=(1,1,i,i)^T/2$ obeying $\mathcal{O}\phi_{y,1}=-\phi_{y,1}$, $\mathcal{O}\phi_{y,2}=\phi_{y,2}$. The perturbation $H_{k_x}$ in the reduced Hilbert space is then $\alpha_x k_x I_{2\times2}+\Delta k_x\sigma_z$. For $\Delta<1$ in the WP phase, the second term lifts the degeneracy of the two edge modes while preserves the chirality; For $\Delta>1$ in the SP phase, the second term dominates, driving the reversal of group velocity for one of the edge modes. Taking together, the edge theory corroborates the Majorana edge modes depicted in Fig. \ref{edge}.

\begin{figure}[t]
\centering
\includegraphics[width=3.25in]{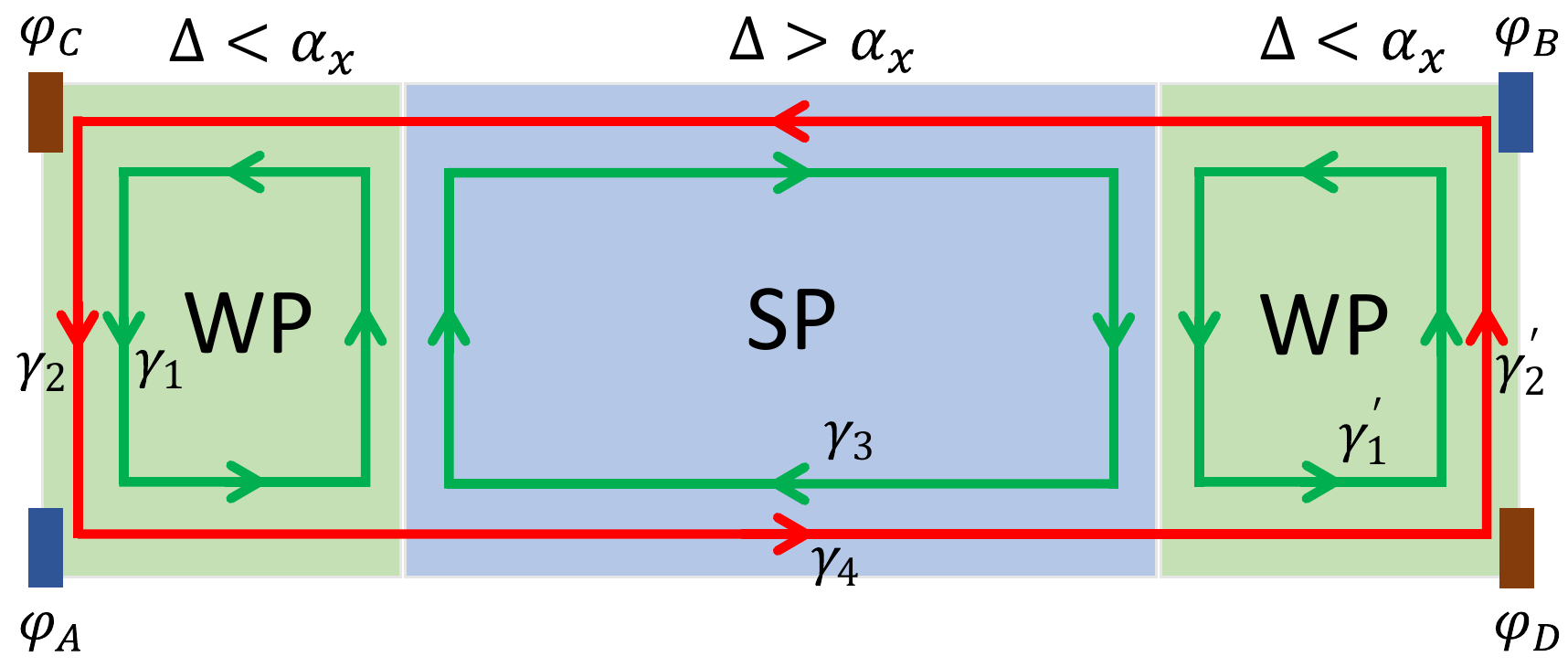}
\caption{The WP-SP-WP topological superconductor junction. For the WP superconductor ($\mathcal{C}=2$) on the two sides, two co-propagating Majorana fermions flow on the edge. For the SP superconductor in the middle, $C_{\mathcal{O}}=1$, a pair of Majorana fermions are counter-propagating. The incoming electrons from leads $A$ and $B$ propagating along the edge are detected in leads $C$ and $D$. Here $\gamma_i~ (i=1,...,4)$ and $\gamma'_{i}~(i=1,2)$ label the corresponding Majorana edge modes.}
\label{fig2}
\end{figure}

{\color{blue}\textit{Braiding.}} Despite the above edge states are not isolated Majorana zero modes or anyons, their propagation in heterostructures can be manipulated to mimic the braiding operation, and further used to implement topologically protected quantum bits and gates. Previously, Ref.~\cite{topoquantumchiral} has demonstrated braiding-like unitary transformation using quantum anomalous Hall insulator (QAHI) and $p_x+ip_y$ chiral topological superconductor, which can be realized for example using a QAHI in proximity with an $s$-wave superconductor. Here, we show that the WP and SP superconductors discussed here can be utilized to form a junction to carry out similar braiding transformations of Majorana fermions. The junction is schematically shown in Fig. \ref{fig2}. In the middle region $\Delta$ is large and the superconductor is in the SP phase. It is sandwiched by two WP regions with smaller $\Delta$. The spatially dependent $\Delta$ can be realized by controlling the proximity coupling strength, for example, by varying the insulating barrier.

From the edge wave functions $\Psi_{j}$ above, one can directly construct $\gamma_j$, the field operator for Majorana fermions. Each edge hosts two Majorana modes. For example, we have $\gamma_1$ and $\gamma_2$ in the WP region and $\gamma_3$ and $\gamma_4$ in the SP region [Fig. \ref{fig2}]. The edge Hamiltonian can be written as, e.g., $H_{edge}(x)=-i\hbar \sum_j v_j \gamma_j(x) \partial_x \gamma_j(x)$ for an edge along $x$. Note that at the interface between WP and SP phases, there are also a pair of co-propagating Majorana modes, in contrast to the proposal in Ref. \cite{topoquantumchiral}. Four external leads, A to D, are attached to the junction. A complex fermion (say an electron) $\varphi_A$ injected from lead $A$ becomes fractionalized, i.e. two Majorana fermions $\gamma_1$ and $\gamma_2$, $\varphi_A=\gamma_1+i\gamma_2$, which follow different propagation paths. For example, $\gamma_2$ will reach lead $D$ while $\gamma_1$ cannot. Similarly, electron from lead $B$ gives rise to Majorana fermion $\gamma'_1$ and $\gamma'_2$. The outgoing electron states in leads $C$ and $D$, which are spatially separated, then become entangled.  As the wave packet of the injected electron evolves, $\gamma_2$ will ditch its original partner $\gamma_1$ and merge with $\gamma'_1$ to form an outgoing electron in lead $D$.  As detailed in Ref. \cite{topoquantumchiral}, using the occupation number $0$ and $1$ as the qubit states for the leads, the unitary evolution involving switching Majorana partners is identical mathematically to braiding. Key to this process is the non-local nature of the edge Majorana fermions.

While the proposal of Ref. \cite{topoquantumchiral} requires all edge states to be chiral, here we explicitly show that this is not necessary. The middle superconductor could be helical, as long as the Majorana edge states are protected by some symmetry, in our case symmetry $\mathcal{O}$, and ensured by bulk topological invariant such as $C_\mathcal{O}$. Our model and analysis suggest that there exist rich families of Majorana fermions associated with additional symmetries beyond the AZ classification, and exploring their properties may prove useful for manipulating Majorana fermions for topological quantum computing.

\begin{acknowledgments}
This research is supported by NSF PHYS-1707484 and AFOSR Grant No. FA9550-16-1-0006.
\end{acknowledgments}

\end{document}